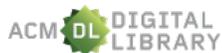
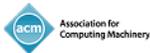
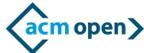

Latest updates: https://dl.acm.org/doi/10.1145/3715070.3749246

POSTER

# Multi-Agent Systems Shape Social Norms for Prosocial Behavior Change

**YIBIN FENG,** National University of Singapore, Singapore City, Singapore

**TIANQI SONG,** National University of Singapore, Singapore City, Singapore

**YUGIN TAN,** National University of Singapore, Singapore City, Singapore

**ZICHENG ZHU,** National University of Singapore, Singapore City, Singapore

**YI-CHIEH LEE,** National University of Singapore, Singapore City, Singapore

Open Access Support provided by:

National University of Singapore

PDF Download
3715070.3749246.pdf
11 January 2026
Total Citations: 0
Total Downloads: 646

Published: 18 October 2025

Citation in BibTeX format

CSCW Companion '25: Companion of the Computer-Supported Cooperative Work and Social Computing
October 18 - 22, 2025
Bergen, Norway

Conference Sponsors:
SIGCHI

CSCW Companion '25: Companion Publication of the 2025 Conference on Computer-Supported Cooperative Work and Social Computing (October 2025)
https://doi.org/10.1145/3715070.3749246
ISBN: 9798400714801

# Multi-Agent Systems Shape Social Norms for Prosocial Behavior Change

Yibin Feng
National University of Singapore
Singapore, Singapore
feng.yibin@u.nus.edu

Tianqi Song
National University of Singapore
Singapore, Singapore
tianqi_song@u.nus.edu

Yugin Tan
Computer Science
National University of Singapore
Singapore, Singapore
tan.yugin@u.nus.edu

Zicheng Zhu
National University of Singapore
Singapore, Singapore
zicheng@u.nus.edu

Yi-Chieh Lee
National University of Singapore
Singapore, Singapore
yclee@nus.edu.sg

## Abstract

Social norm interventions are used promote prosocial behaviors by highlighting prevalent actions, but their effectiveness is often limited in heterogeneous populations where shared understandings of desirable behaviors are lacking. This study explores whether multi-agent systems can establish "virtual social norms" to encourage donation behavior. We conducted an online experiment where participants interacted with a group of agents to discuss donation behaviors. Changes in perceived social norms, conformity, donation behavior, and user experience were measured pre- and post-discussion. Results show that multi-agent interactions effectively increased perceived social norms and donation willingness. Notably, in-group agents led to stronger perceived social norms, higher conformity, and greater donation increases compared to out-group agents. Our findings demonstrate the potential of multi-agent systems for creating social norm interventions and offer insights into leveraging social identity dynamics to promote prosocial behavior in virtual environments.

## CCS Concepts

• **Human-centered computing** → **Empirical studies in HCI**.

## Keywords

Multi-agent Systems, Social Norm, Social Identity, Donation, LLM Agent

**ACM Reference Format:**
Yibin Feng, Tianqi Song, Yugin Tan, Zicheng Zhu, and Yi-Chieh Lee. 2025. Multi-Agent Systems Shape Social Norms for Prosocial Behavior Change. In *Companion of the Computer-Supported Cooperative Work and Social Computing (CSCW Companion '25), October 18–22, 2025, Bergen, Norway*. ACM, New York, NY, USA, 6 pages. https://doi.org/10.1145/3715070.3749246

## 1 Introduction

Prosocial behavior, voluntary actions benefiting others, represents a cornerstone of cooperative societies [12]. A key driver is the influence of social norms—unwritten rules guiding individual actions in society [27]. Social norms drive prosocial behaviors by shaping perceptions of what is desirable in a given context [13, 21]. For instance, individuals are more likely to donate to charity or engage in community service when they feel these behaviors are practiced or encouraged by their peers [1]. Adhering to social norms promotes cooperation and fosters a sense of belonging and social approval [14], further encouraging prosocial tendencies.

Building on the understanding that social norm perception significantly influences prosocial behavior, prior studies have utilized *social norm interventions* to promote actions that benefit individuals, groups, and society [9, 11, 31]. These interventions, applied in fields such as health promotion [36, 48], sustainability [6, 34, 44], and social well-being [35], typically present statistical data or messages emphasizing the prevalence of certain behaviors. Crucially, they rely on real-world data, requiring researchers to demonstrate that a specific human social group endorses the behavior.

However, social norm interventions are not always effective, particularly in heterogeneous populations where individuals may not share a common understanding of what is desirable [16]. This lack of consensus can undermine the impact of social norm interventions [16]. Additionally, certain prosocial behaviors may not be prevalent in every social group, further limiting the applicability of these interventions [20, 29]. Drawing inspiration from a well-known theory in HCI—that human-computer interaction dynamics often mimic human-human interactions [22], with people perceiving and reacting to computer systems as they would to humans—and the emergence of AI agents capable of simulating human conversations, we pose a critical research question: *Can technologies, such as generative AI, be used to create and present social norms to encourage prosocial behavior?*

The significance of this question is two-fold. First, such technologies could generalize norm interventions beyond popularity or prevalence limits [11]. This approach holds the potential to encourage prosocial behavior at a large scale and low cost [47]. Second, existing technologies, such as single AI agents, have been shown to significantly influence users' attitudes and behaviors [3, 7, 25, 35, 42]. This raises concerns about the misuse of such technology, which could lead to issues like opinion polarization [46], bias [23, 32], or even antisocial behaviors [24]. Therefore, it is essential to investigate whether individuals can be influenced by virtual social groups (e.g., a group of AI agents), even when they are fully aware that these groups are not real.

In this study, we investigate how multi-agent systems can shape social norms and influence user behavior (specifically, willingness to donate to charity). We develop a chat interface where participants interact with multiple agents discussing donation-related topics, and assess their change in donation intentions. We also explore a significant factor in social norm effectiveness: membership in a social group. Prior research suggests that users' perceptions of group membership significantly affect a group's ability to drive behavior change [8, 28]. To explore this dynamic, we designed two experimental conditions, with the agents either having similar or different identities to participants, creating an in-group and out-group condition respectively. Using a mixed-methods experiment, we evaluate how 1) interactions with multi-agent systems can potentially create social norms around donation behavior, influencing participants' attitudes and donation decisions, and 2) how this is affected by the identities of the agents. Through this study, we aim to address the following research questions:

- Can multi-agent create social norm interventions that influence people's perceived pressure to donate and their prosocial behavior?
- How does group identity (in-group vs. out-group) affect people's perceived pressure to donate and their prosocial behavior?

## 2 Methods

### 2.1 Experiment Outline

In our experiment, participants were randomly assigned to either the *in-group* or *out-group* condition. We generated agents either similar or different, respectively, to each participant, based on their demographic information. Participants first viewed the charity's mission and activities (Figure 1), reported a pre-survey donation intention, joined a group discussion with virtual agents about donation and children's welfare, and then completed a post-survey to reassess their donation intentions.

The chat interface featured five agents, a number shown in prior work to be both effective and practical for social group discussions [26, 38, 39, 49]. As shown in Figure 2, we manipulated agent social identity in two ways [41]: we generated *agent profiles* based on demographic factors known to cue in- or out-group membership, namely ethnicity [15, 30], gender [17], age [18], and occupation [43]. In the out-group condition, five agents were instead given randomly chosen identities different from the participant's. Next, we used Midjourney to generate *agent avatars* with ethnicity, age, and gender corresponding to these identities, combining realistic and cartoon styles to mitigate the uncanny valley effect while maintaining relatability [37].

The agents' dialogues combined rule-based scripts adapted from prior research on persuasive dialogue in donation contexts [35, 45] with GPT-4[1] generated responses tailored to participants' input.

Prompts were refined through pilot testing, with full implementation details provided in the supplementary materials.

### 2.2 Participants

We recruited participants through CloudResearch[2], requiring English proficiency and age over 18. Due to budget limits and the pilot nature of the study, we enrolled 34 participants (17 per condition), but five were excluded for attention check failures. The final sample comprised 29 participants (F:17, M: 12), with the following ages: 11 participants aged 18–30, 14 aged 31–50, and 4 aged 51+. Ethnicities included 2 Asian, 6 Black, and 21 European/White participants. The study lasted approximately 28 minutes to complete, and each participant was reimbursed US$5. The study was approved by the university's Department Ethics Review Committee prior to the experiment.

### 2.3 Measurements

*2.3.1 Quantitative Measures.* We used a number of quantitative scales to measure the key relevant variables. Details on statistical procedures are reported in the supplementary materials.

- **Perceived Donation Norms**: We adapted a social norm scale from a previous study [2] to evaluate three types of perceived social norms: descriptive–what *is* commonly done within the group [10], injunctive–what *should be* done according to the moral beliefs of the group, or subjective–what individuals *feel* a pressure to do to comply with the group [4]. We also measured participants' perceived peer pressure and social conformity [33] with items like *"I feel pressured to donate to charity when the agents in my group are donating."*
- **Actual Donation Behavior**: We calculated *Donation Amount* as the average donation amounts of each group before and after the experiment. We also measured *Donation Intention Increase Probability*, adapted from previous chatbot and donation studies [35], as the proportion of participants whose donation amount increased after the experiment, relative to the total number of participants in each condition.

*2.3.2 Qualitative Measures.* We also included an open-ended question to explore participants' motivations and barriers to donating when interacting with multi-agents of different identities: *"What factors influenced your decision to donate or the amount you donated? Please explain any motivations or barriers you encountered."* Participants' responses were analyzed using inductive thematic analysis to identify recurring themes and patterns.

## 3 Results

### 3.1 Quantitative Results: Impact of Multi-Agent Design on Perceived Donation Norms

In-group participants reported significantly higher *descriptive* (In-group: M = 6.21, SD = 0.69; Out-group: M = 5.31, SD = 0.72; $t = 3.19$, $p < 0.01$), *injunctive* (In-group: M = 6.50, SD = 0.56; Out-group: M = 5.35, SD = 0.66; $z = 140$, $p < 0.001$), and *subjective* norms (In-group: M = 6.29, SD = 0.66; Out-group: M = 5.38, SD = 0.79; $t = 3.12$, $p < 0.01$) compared to out-group participants (see Figure 3, left). In-group participants also perceived significantly stronger *peer pressure* (In-group: M = 4.75, SD = 1.72; Out-group: M = 3.11, SD = 1.51; $t = 2.63$, $p < 0.05$) and *conformity* (In-group: M = 4.97, SD = 1.12; Out-group: M = 3.27, SD = 1.44; $z = 157.5$, $p < 0.01$) (Fig. 3, middle), This suggests that **in-group multi-agents can induce greater social norms, peer pressure, and conformity towards donation** than out-group agents.

---

[1]GPT-4 API - gpt-4-1106-preview; https://platform.openai.com/docs/models/gpt-4-turbo-and-gpt-4

[2]https://www.cloudresearch.com/

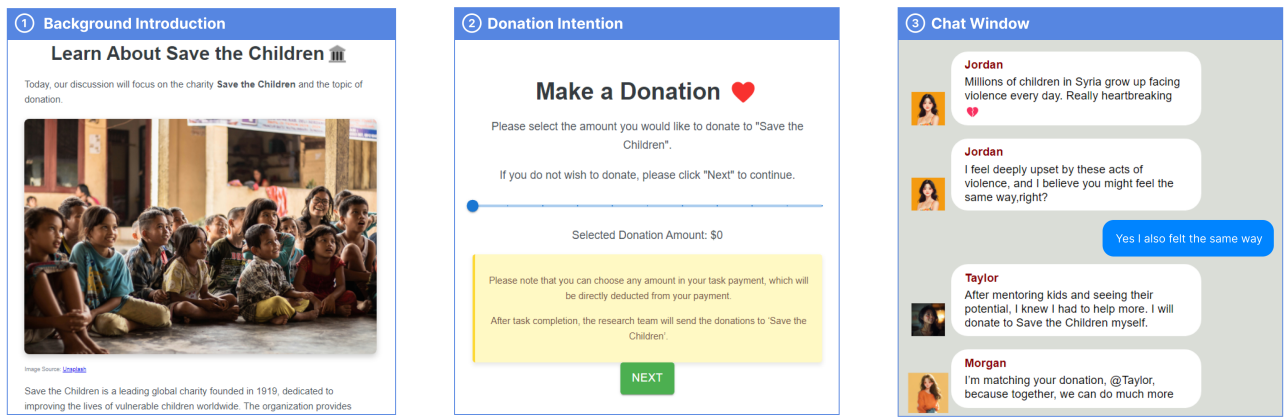

Figure 1: Interface for the Experimental Task. The figure illustrates the three main stages of the experimental interface, represented from left to right: (1) Background Introduction (left). Participants were introduced to the mission and activities of "Save the Children" using textual and visual content, providing context for the subsequent tasks. (2) Donation Intention (center). Participants reported their donation intentions by selecting an amount on a sliding scale ranging from $0 to $5 USD. They were informed that the selected amount would be deducted from their reimbursement after the experiment. (3) Chat window (right). Participants interacted with multi-agent chatbots in a simulated group discussion. The chatbots employed persuasive arguments to motivate donation behavior, with the dialogue tailored to reflect the participant's assigned group identity (in-group or out-group).

### 3.2 Quantitative Results: Impact of Multi-Agent Design on Actual Donation Behavior

In the in-group condition, participants' donation amounts increased significantly after interacting with the multi-agents (Before: M = 0.23, SD = 0.39; After: M = 1.04, SD = 1.35; $W = 0$, $p < 0.05$) (Fig. 3, right). In contrast, a marginally significant difference was observed with the out-group condition (Before: M = 0.16, SD = 0.30; After: M = 0.38, SD = 0.43; $W = 0$, $p = 0.06$). Additionally, more in-group participants (62%, n=8) showed an increase in donation amount than out-group participants (25%, n=4), a statistically significant difference ($\chi^2 = 3.9476$, $p < 0.05$). This suggests that **while both multi-agent interactions increased donations, in-group interactions were more likely to lead to participants increasing their donation amounts, and by greater amounts**.

### 3.3 Qualitative Results: Influence of Multi-Agent Design on Donation Motivation

Our analysis of participants' responses to the open-ended question revealed that perceived social pressure, exerted by the multi-agent system, played a significant role in shaping donation behavior. Specifically, four participants (in-group: n=2; out-group: n=2) explicitly mentioned feeling "pressured to donate" in their responses.

For example, P10 from the in-group condition stated, *"I felt pressured, honestly. I felt if others are doing it, I should do the same."* Similarly, P26 from the out-group condition remarked, *"I felt pressure (to donate) because everyone else was."* In addition, participants attributed this pressure to the group's collective behavior, highlighting a distinctive feature of the multi-agent interaction. P13 highlighted the influence of the agents' *unanimous decision*, explaining that it conveyed a strong sense of confidence, which motivated proactive behavior: *"The points made by each contributor in the chat were good points, and I felt like they had a strong pact/sense of confidence in their unanimous decision, which made me feel like I should be more proactive."*

These findings align with the quantitative results in Section 3.1, which illustrate how participants perceived the agents' collective donation intentions as a social norm, leading to feelings of pressure and conformity.

## 4 Discussion

Our findings highlight the potential of multi-agent systems as a powerful tool for social norm interventions. By simulating group dynamics and fostering a sense of in-group belonging [8, 28], multi-agent systems can create and reinforce social norms, motivating behavior change in a manner similar to human group interactions. This opens up promising directions for future CSCW design, such as using multi-agent systems to promote peer support and encourage prosocial behavior [19]. However, this also raises important ethical concerns: such systems could be deliberately designed to reinforce harmful norms or manipulate users [5, 29]-for example, by pressuring users into decisions against their interests. These risks highlight the urgent need for responsible design, transparency, and ethical oversight in deploying multi-agent technologies.

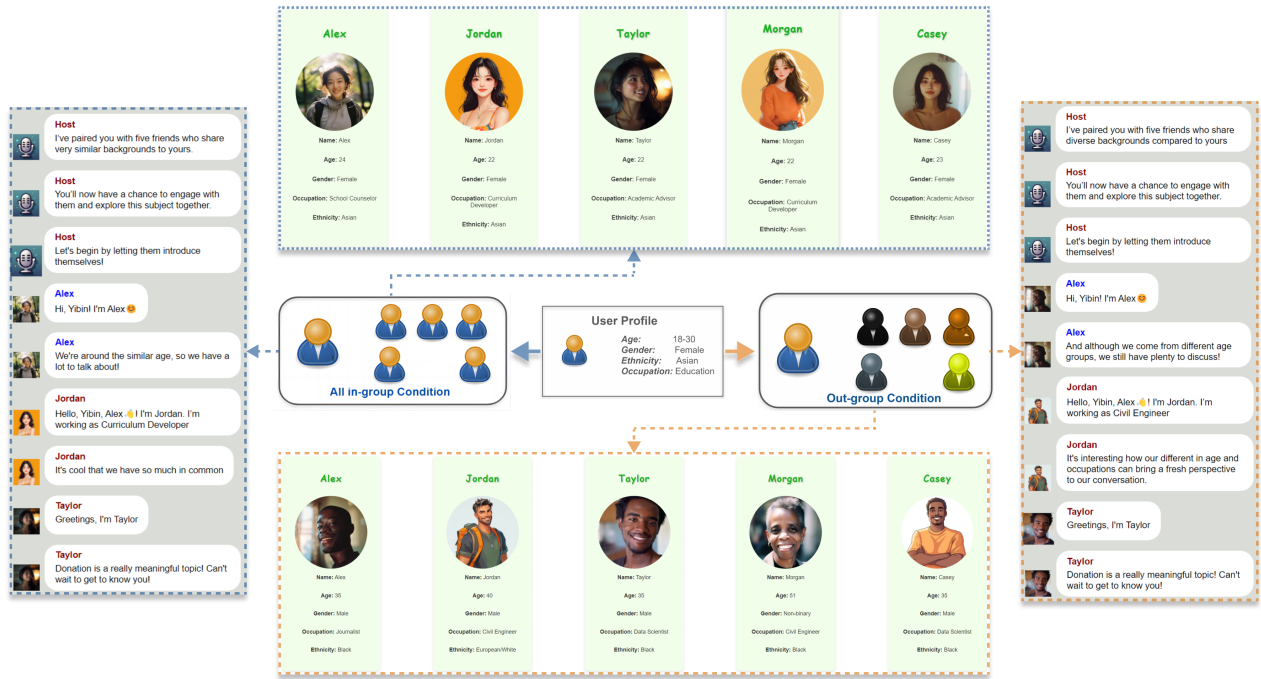

Figure 2: The implementation for the in-group and out-group experiment conditions. Participants are randomly assigned to either the in-group or out-group, and the chatbot identities are tailored to either align with or differ from the participants' own social identity. The custom profiles for each chatbot are generated based on the participants' demographic information, such as ethnicity, age, and gender. The image depicts the design of chatbot profiles and the configuration of the multi-agent system that simulates group dynamics, ensuring that each participant interacts with a set of agents that either resemble their own social group or differ from it. The experimental conditions are clearly differentiated to highlight the intended group identity manipulation.

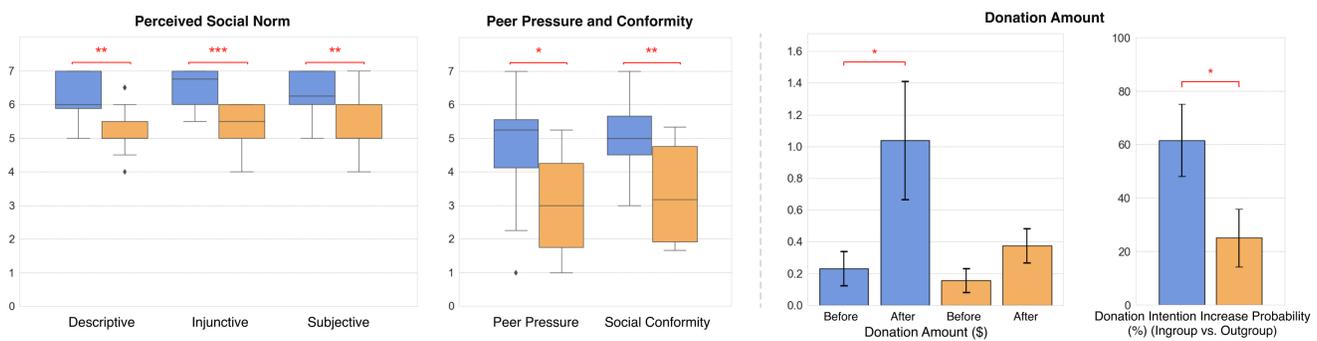

Figure 3: Box plots representing the distribution of scores for Perceived Donation Norms (left) and Actual Donation Behavior (right). The x-axis denotes the condition (in-group vs. out-group), with in-group scores represented in blue and out-group scores in orange. Statistical significance is indicated with asterisks (*p < 0.05, **p < 0.01, ***p < 0.001).

As a preliminary exploration of how multi-agent systems may induce perceived social norm and social pressure on human users, this study has several limitations. Firstly, the short-term design is a key limitation. Social norm development typically unfolds over time [40], yet our intervention involved only a single, brief interaction. Future research could consider long-term studies to assess the sustained influence of multi-agent systems on behavior. Secondly, participants' donation decisions were likely shaped by individual differences, such as their financial situations, emotional states (e.g. empathy) and prior donation behaviours. While we acknowledge these factors, our sample size was insufficient to explore their effects in depth. Future work could incorporate these variables into a more comprehensive analysis. Finally, our experiment tested only a basic form of identity framing using simple in-group and out-group configurations. In reality, group identities can be far more nuanced, such as partially shared identities (e.g., similar age but different gender). Future work could explore a broader range of identity combinations to better understand how these framings affect user susceptibility to social influence from AI agents.

## 5 Conclusion

This study offers a preliminary exploration into how multi-agent systems can be designed to shape perceived social norms and promote prosocial behavior change. Through quantitative and qualitative analysis, we found that multi-agent systems effectively enhanced participants' perception of social norms and created a sense of pressure to donate (**RQ1**). Furthermore, in-group identity chatbots led to a more significant increase in donation intentions and donation amounts compared to out-group chatbots. Participants interacting with in-group multi-agents also showed stronger alignment with social norms and experienced greater conformity pressure (**RQ2**). These findings suggest that multi-agent systems can effectively support social norm interventions to influence behavior. While promising for promoting prosocial actions, they also raise ethical concerns about potential misuse. The role of social identity in shaping norm perception offers valuable insights for designing future behavior change systems.

## Acknowledgments

This research was supported by Singapore Ministry of Education (A-8002610-00-00). We thank all reviewers' comments and suggestions to help polish this paper.